\documentclass[9pt,twocolumn,twoside]{pnas-new}
\usepackage{bm}
\usepackage{bbm}
\templatetype{pnasresearcharticle} 

\title{Market structure dynamics during COVID-19 outbreak}

\author[]{Pier Francesco Procacci}
\author[]{Carolyn E. Phelan} 
\author[]{Tomaso Aste} 
\affil[]{Department of Computer Science, UCL, Gower Street, WC1E6BT London, UK }

\begin{abstract}
In this note we discuss the impact of the COVID-19 outbreak from the perspective of market-structure. 
We observe that US market-structure has dramatically changed during the past four weeks and that the level of change has followed the number of infected cases reported in the USA. 
Presently, market-structure resembles most closely the structure during the middle of the 2008 crisis but there are signs that it may be starting to evolve into a new structure altogether. 
This is the first article of a series where we will be analysing and discussing market-structure as it evolves to a state of further instability or, more optimistically, stabilisation and recovery.
\end{abstract}

\dates{\today}

\begin{document}

\maketitle

\ifthenelse{\boolean{shortarticle}}{\ifthenelse{\boolean{singlecolumn}}{\abscontentformatted}{\abscontent}}{}
\section*{\vskip-1.3cm Motivations}\vskip-0.2cm
COVID-19 outbreak is an unprecedented event in modern human history with potentially catastrophic human consequences. 
While at the moment the focus is on saving human lives and preventing the virus spreading further, this pandemic is also having profound effects on society, the economy and the financial system. 
Markets have seen significant numbers of investors  selling off and rebalancing their portfolios with less risky assets. This has resulted in large losses and high volatilities, typical of crisis periods. 
We aim to better understand present market dynamics with the scope of the containment and possible prevention of future periods of extreme market instability and their consequent effects on the economy and people's lives.

\section*{\vskip-0.7cm Methods}\vskip-0.2cm
We used the unsupervised clustering methodology described in \cite{procacci2019forecasting} to automatically extract four inherent market-structures associated with a set of 623 equities continuously traded in the US market during the period from February 1999 to March 20, 2020.  
The clustering was performed by maximising the following adjusted log-likelihood:
\begin{align} \nonumber
  \mathcal{\tilde L}_{t,k} \!\!=\! -\frac{1}{2}\!(\bm{X}_t\!-\!\bm{\mu}_k)^T\!\! \bm{J}_k\! \;(\bm{X}_t\!-\!\bm{\mu}_k) \!+\! \frac{1}{2} \log\; \lvert \bm{J}_k\rvert 
  \!-\! \gamma \mathbbm{1} \lbrace \mathcal K_{t-1}\!\not = k \rbrace .
  \label{likelihood_def0}
\end{align}
where $\bm{X}_t\in \mathbb R^{n,1}$  is the vector of log-returns at time $t$; $\bm{\mu}_k \in \mathbb R^{n,1}$ is the vector of the expected values for cluster $k$; $\bm{J}_k\in \mathbb R^{n,n}$ is the sparse precision matrix for cluster $k$ computed via the LoGo method (see \citep{AsteTMFG,Aste_Parsimonious}); $\gamma$ is a parameter penalizing state switching. In the present analysis we use $\gamma = 100$, but results are consistent across a large range of values of this parameter. 
Note that the present approach is slightly different from  \cite{procacci2019forecasting} where the Mahalanobis distance was minimized instead. 
It was indeed noticed already in \cite{procacci2019forecasting} that clustering through likelihood maximisation is more efficient in detecting states of market stress. 
The results reported in this note refer to the case of four clusters but the outcomes are robust with respect to the number of clusters and analogous results can be obtained for two or six clusters as well.  

\section*{\vskip-0.8cm Results}
\begin{figure}[]
\centering
\vskip-3cm
\includegraphics[width=.8\linewidth]{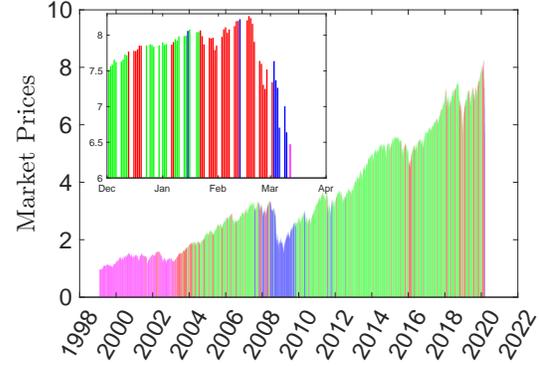}
\caption{Market states during the period 02/1999-03/2020. 
The y-axis reports and average market price and the color of the bars correspond to the market-state assigned by the unsupervised clustering procedure introduced in \cite{procacci2019forecasting}.}
\label{fig:MarketStates}
\end{figure}

\begin{figure}[]
\centering
\vskip-0.2cm
\includegraphics[width=.8\linewidth]{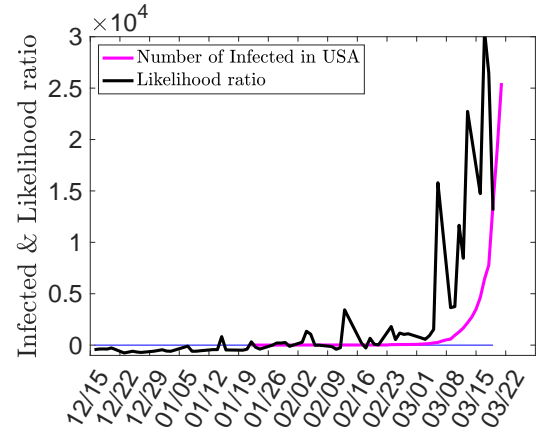}
\caption{Comparison between the number of COVID-19 infected cases in US and the logarithm of the ratio between the likelihood of the bull-state (green bars in Fig.\ref{fig:MarketStates}) and the stress-state  (blue bars in Fig.\ref{fig:MarketStates}). 
The period is from December 15 2019 to March 20 2020.}
\label{fig:Likelihood}
\vskip-0.5cm
\end{figure}

\vskip-0.2cm
The  clustering structure is shown in Fig.\ref{fig:MarketStates} where a mean market price is reported with bars whose colour represents the cluster associated to each day. 
Note the central part of the 2008 crisis is associated with a state (blue bars) that has again become prevalent during the last few weeks (see inset).  
We compare the likelihood of this `crisis' state with the likelihood associated with the state which is instead prevalent during the long `bull' period post 2008 (green bars). 
The result is shown in Fig.\ref{fig:Likelihood} where the logarithm of the ratio between the likelihoods of the crisis and bull states is reported.
We note that the bull-state prevails until February 2020 producing a negative log-ratio, afterwards the crisis-state becomes more representative and eventually becomes extremely dominant in March. 
The timing of the surge in the dominance of the crisis-state is consistent with that of the surge of US confirmed COVID-19 cases. 
It must be pointed out that this is an initial analysis and developing a definitive representation is hampered by a current lack of observations. However our very latest analysis, which is still in the initial stage, suggests the present market state may ultimately be classified as distinct from that of the 2008 crisis with some similarities with the late 90' states. We will publish further results as they become available.

%
\section*{\vskip-0.8cm References}

\end{document}